\input{aipcheck}

\documentclass[
    ,final            
  ]
  {aipproc}

\layoutstyle{6x9}

\newcommand{\ltsima}{$\; \buildrel < \over \sim \;$}
\newcommand{\simlt}{\lower.5ex\hbox{\ltsima}}
\newcommand{\gtsima}{$\; \buildrel > \over \sim \;$}
\newcommand{\simgt}{\lower.5ex\hbox{\gtsima}}

\newcommand{\cgs}{ ${\rm erg~cm}^{-2}~{\rm s}^{-1}$} 
\newcommand{\lum}{\rm erg~s$^{-1}$}

\def\lesssim{\mathrel{\hbox{\rlap{\hbox{\lower4pt\hbox{$\sim$}}}\hbox{$<$}}}}
\def\gtrsim{\mathrel{\hbox{\rlap{\hbox{\lower4pt\hbox{$\sim$}}}\hbox{$>$}}}}

\def\micron{\hbox{$\mu$m}}
\def\mujy{\hbox{$\mu$Jy}}

\def\ab1450{$AB_{1450(1+z)}$}

\def\xray{\hbox{X-ray}}

\def\ks{$K_{\rm s}$}

\def\lbol{\hbox{$L_{\rm bol}$}}
\def\kbol{\hbox{$k_{\rm bol}$}}
\def\ledd{\hbox{$L_{\rm Edd}$}}
\def\mstar{\hbox{$M_{\star}$}}
\def\mbh{\hbox{$M_{\rm BH}$}}

\def\msun{M$_{\odot}$}
\def\edd_ratio{$\log\ L_{\rm bol}/L_{\rm Edd}$}
\providecommand{\eprint}[2][]{\url{#2}}

\def\chandra{{\it Chandra\/}}

\def\heao1{{\it HEAO-1\/}}

\def\spitzer{{\it Spitzer\/}}

\def\xmm{{XMM-{\it Newton\/}}}

\def\aj{AJ}
\def\araa{ARA\&A}
\def\apj{ApJ}
\def\apjl{ApJ}
\def\apjs{ApJS}

\def\aap{A\&A}

\def\mnras{MNRAS}

\begin{document}

\title{The obscured \xray\ source population in the HELLAS2XMM survey: the \spitzer\ view}

\classification{98.54.-h, 98.58.Jg}
\keywords      {Galaxies: active --
                Galaxies: nuclei --
                X-rays: galaxies}

\author{Cristian Vignali}{
  address={Dipartimento di Astronomia, Universit\`a di Bologna, 
Via Ranzani 1, I--40127 Bologna, Italy}
,altaddress={INAF -- Osservatorio Astronomico di Bologna, Via Ranzani 1, 
I--40127 Bologna, Italy}
}

\author{Francesca Pozzi}{
  address={Dipartimento di Astronomia, Universit\`a di Bologna, 
Via Ranzani 1, I--40127 Bologna, Italy}
}

\author{Andrea Comastri}{
  address={INAF -- Osservatorio Astronomico di Bologna, Via Ranzani 1, 
I--40127 Bologna, Italy}
}

\author{Lucia Pozzetti}{
  address={INAF -- Osservatorio Astronomico di Bologna, Via Ranzani 1, 
I--40127 Bologna, Italy}
}

\author{Marco Mignoli}{
  address={INAF -- Osservatorio Astronomico di Bologna, Via Ranzani 1, 
I--40127 Bologna, Italy}
}

\author{Carlotta Gruppioni}{
  address={INAF -- Osservatorio Astronomico di Bologna, Via Ranzani 1, 
I--40127 Bologna, Italy}
}

\author{Giovanni Zamorani}{
  address={INAF -- Osservatorio Astronomico di Bologna, Via Ranzani 1, 
I--40127 Bologna, Italy}
}

\author{Carlo Lari}{
  address={INAF -- Istituto di Radioastronomia (IRA), Via Gobetti 101, I--40129 Bologna, Italy}
}

\author{Francesca Civano}{
  address={Dipartimento di Astronomia, Universit\`a di Bologna, 
Via Ranzani 1, I--40127 Bologna, Italy}
}

\author{Marcella Brusa}{
  address={Max Planck Institut f\"ur Extraterrestrische Physik (MPE), Giessenbachstrasse 1, 
D--85748 Garching bei M\"unchen, Germany}
}

\author{Fabrizio Fiore}{
  address={INAF -- Osservatorio Astronomico di Roma, Via Frascati 33, I--00040
Monteporzio-Catone (RM), Italy}
}

\author{Roberto Maiolino}{
  address={INAF -- Osservatorio Astronomico di Roma, Via Frascati 33, I--00040
Monteporzio-Catone (RM), Italy}
}

\begin{abstract}
Recent \xray\ surveys have provided a large number of high-luminosity, obscured Active Galactic 
Nuclei (AGN), the so-called Type~2 quasars. Despite the large amount of multi-wavelength supporting 
data, the main parameters related to the black holes harbored in such AGN are still poorly 
known. Here we present the preliminary results obtained for a sample of eight Type~2 quasars 
in the redshift range $\approx$~0.9--2.1 selected from the HELLAS2XMM survey, 
for which we used \ks-band, \spitzer\ IRAC and MIPS data at 24~\micron\  
to estimate bolometric corrections, black hole masses, and Eddington ratios. 
\end{abstract}

\maketitle


\section{Introduction}
Over the last six years, the \xray\ surveys carried out by \chandra\ and \xmm\ 
(e.g., \citep{gia02,ale03,fio03}; see \citep{bh05} for a review) 
have provided remarkable results in resolving a significant fraction of the 
cosmic \xray\ background (XRB; \citep{com95,gil07}), 
up to $\approx$~80\% in the 2--8~keV band (e.g., \citep{feb04,hm06}). 
Despite the idea that a large fraction of the accretion-driven energy density in the Universe 
resides in obscured \xray\ sources has been widely supported and accepted 
(e.g., \citep{bar05,hop06}), until recently only limited information was available to 
properly characterize the broad-band emission of the counterparts of the \xray\ obscured sources 
and provide a reliable estimate of their bolometric output. 

In this context, \spitzer\ data have provided a major step forward the understanding of the 
broad-band properties of the \xray\ source populations. If, on one hand, \spitzer\ data have 
allowed to pursue the ``pioneering'' studies of \cite{elv94} on the spectral energy 
distributions (SEDs) of broad-line (Type~1), unobscured quasars at higher redshifts 
(e.g., \citep{ric06}), on the other hand they have produced significant results in the 
definition of the SEDs of narrow-line (Type~2), obscured AGN (e.g., \citep{pol06}). 

In this work we aim at providing a robust determination of the bolometric luminosity for hard 
\xray\ selected obscured AGN. This result can be achieved by effectively disentangling 
the nuclear emission related to the active nucleus from the host galaxy starlight, which represents 
the dominant component (at least for our obscured sources) at optical and near-infrared (near-IR) 
wavelengths.

\section{Sample selection and \ks-band properties}
The sources presented in this work were selected from the HELLAS2XMM survey (\citep{fio03}) 
which, at the 2--10~keV flux limit of $\approx$~10$^{-14}$~\cgs, covers $\approx$~1.4 
square degrees of the sky using \xmm\ archival pointings (\citep{bal02}). 
Approximately 80\% of the HELLAS2XMM sources have a spectroscopic optical classification in the 
final source catalog (see \citep{coc07} for details). 
In particular, we selected eight sources from the original sample of \cite{mig04} which 
are characterized by faint (23.7--25.1) $R$-band magnitudes and bright \ks-band counterparts 
($\approx$~17.6--19.1); all of our sources are therefore classified as extremely red objects 
(EROs, $R-K_{\rm s}>5$ in Vega magnitudes). 
From the good-quality \ks-band images, \cite{mig04} were able to study the 
surface brightness profiles of these sources, obtaining a morphological classification. 
While two sources are associated with point-like objects, the remaining six sources are extended, 
showing a profile typical of elliptical galaxies. 
In this latter class of sources, the active nucleus, although evident in the \xray\ band, 
appears hidden or suppressed at optical and near-IR wavelengths, 
where the observed emission is clearly dominated by the host galaxy starlight. 
The relatively good constraints on the nuclear emission 
in the near-IR represent a starting point for the analysis of the \spitzer\ IRAC and 
MIPS data. 

Due to the faint $R$-band magnitudes of our sources, optical spectroscopy was not feasible 
even with the 8-m telescope facilities; however, the bright near-IR counterparts of our 
sources allowed us to obtain spectroscopic redshifts in the \ks\ band with {\sc ISAAC} 
at {\sc VLT} for two sources: 
one point-like AGN is classified as a Type~1.9 quasar at a redshift of 2.09, while one extended 
source has line ratios typical of a LINER at $z$=1.35 (see \citep{mai06} for further 
details on these classifications). 
For the remaining sources, the redshift has been estimated using the optical and near-IR 
magnitudes, along with the morphological information, as extensively described in $\S$5.1 of 
\cite{mig04}; all of the redshifts are in the range $\approx$~0.9--2.1. 
The large column densities [$\approx10^{22}$ -- a~few$\times10^{23}$~cm$^{-2}$] and the 
\hbox{2--10~keV} 
luminosities [$\approx(1-8)\times10^{44}$~\lum, once corrected for the absorption] place our sources 
in the class of the high-luminosity, obscured AGN, the so-called Type~2 quasars (see, e.g., 
\citep{vig06} and references therein).

\section{\spitzer\ data} 
For our sample of eight sources, we obtained IRAC observations of 480~s integration time and MIPS 
observations at 24~\micron\ for a total integration time per position of $\approx$~1400~s. 
All of the sources are detected in the four IRAC bands and in MIPS; 
the faintest source in MIPS has a 24~\micron\ flux density of $\approx$~150~\mujy\  
($\approx$~5$\sigma$ detection; see \cite{poz07} for further details on data reduction and 
cleaning procedures).

\section{Analysis of the Type~2 quasar spectral energy distributions}
A reliable determination of the bolometric output of our AGN sample requires that the nuclear 
component, directly related to the accretion processes, is disentangled from the emission of the 
host galaxy, which provides a dominant contribution in the optical and \ks\ bands (in the 
case of extended sources, see \citep{mig04}). 
To achieve this goal, we constructed SEDs for all our sources over the optical, near- and 
mid-IR range. At the same time, we used \spitzer\ data to improve our previous estimates
on the source redshift when possible. 
In the following, we consider the sample of six extended sources and two point-like objects 
separately, since a different approach has been adopted for the two sub-samples. 

\subsection{Extended sources}
As already pointed out, from the \ks-band morphological analysis carried out by 
\cite{mig04}, we know that at least up to 2.2~\micron\ (observed frame) the stellar contribution 
is mostly responsible for the emission of these sources. 
At longer wavelengths, the emission of the active nucleus is expected to arise as reprocessed 
radiation of the primary emission, while the emission from the galaxy should drop significantly, 
assuming reasonable elliptical templates. 
Although many models have been developed in the past to deal with circum-nuclear dust emission 
(including the effects of the torus geometry and opening angle, grain size distribution and 
density), in our study we adopted a more phenomenological approach. 
To reproduce the observed data, we used a combination of two components, one for the host galaxy 
and another related to the reprocessing of the nuclear emission. 

For the galaxy component, we adopted a set of early-type galaxy templates obtained from the 
synthetic spectra of \cite{bc03}, assuming a simple stellar population spanning a large range of 
ages (see \citep{poz07} for details). For the nuclear component, we adopted the templates of 
\cite{sil04}, which are based on the interpolation of the observed nuclear IR data 
(at least, up to $\approx$~20~\micron) of a sample of local AGN 
through the radiative transfer models of \cite{gd94}. 
The strength of such an approach is that the nuclear templates depend upon two quantities, 
the intrinsic 2--10~keV luminosity (which provides the normalization of the SED) and the column 
density (responsible for the shape of the SED), and these are known directly from the \xray\ 
spectra (\citep{per04}), once the redshift is known. 

We also used all the available information, extended over the \spitzer\ wavelength range, to 
place better constraints on the source redshift than those reported in \cite{mig04}. 
Overall, we find a good agreement with the redshifts presented in \cite{mig04}, although \spitzer\ 
allows us to provide estimates with lower uncertainties; 
only for one source the redshift is significantly lower ($z\approx$~1 instead of $\approx$~2) 
and likely more reliable. 

\begin{figure}
\includegraphics[height=0.28\textheight]{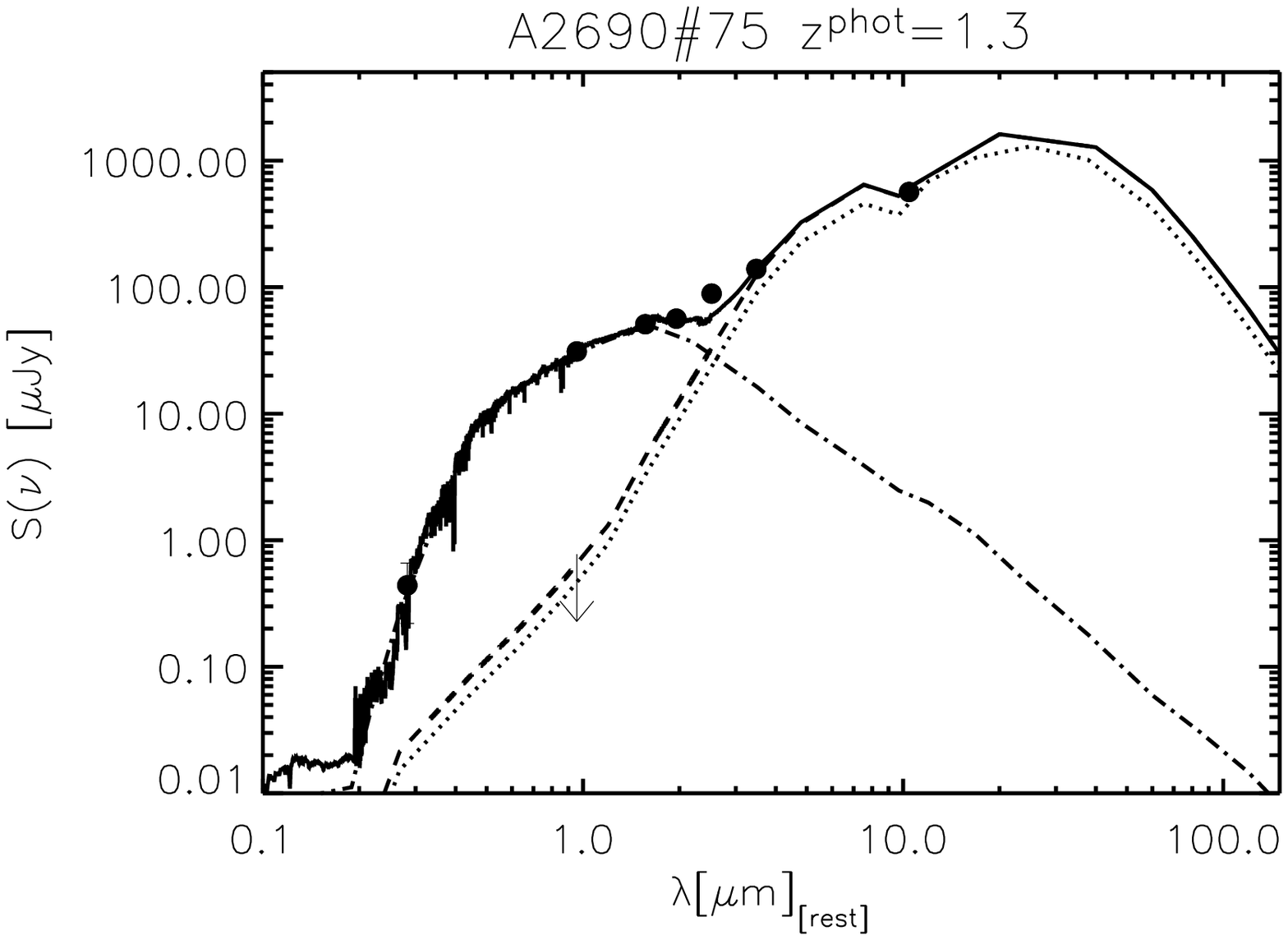}
\includegraphics[height=0.28\textheight]{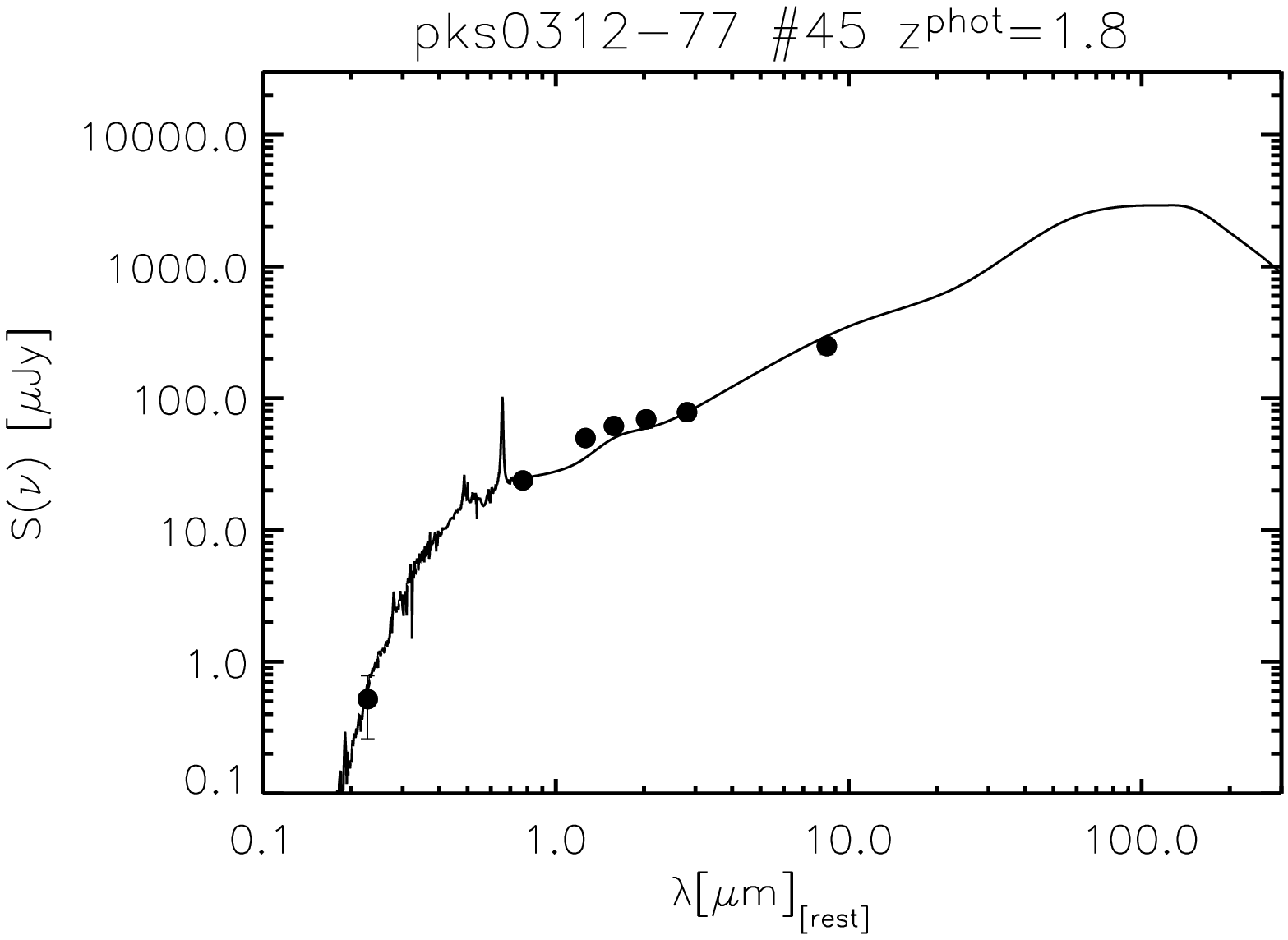}
\caption{\label{seds} Rest-frame SEDs for two representative Type~2 quasars of the current 
sample: an obscured AGN hosted by an elliptical galaxy (on the left) and a point-like AGN 
(on the right). 
{\it (Left)} The observed data (filled circles) are reproduced by summing up (solid line) 
the contribution of an early-type galaxy template (dot-dashed line) to the reprocessed 
nuclear component (dashed line). 
The dotted line shows the nuclear component obtained from the templates of \cite{sil04}, 
normalized using the \xray\ luminosity and column density (i.e., without fitting the data; see text 
and \citep{poz07} for details). 
The downward-pointing arrow indicates the constraint on the nuclear emission derived from the 
\ks-band data (\citep{mig04}). The combination of the two templates is also consistent with the 
$R$-$K_{\rm s}$ color. 
{\it (Right)} The observed data (filled circles) are well reproduced by the red quasar template 
from \cite{pol06} (solid line). 
}
\end{figure}

The data are well reproduced by the sum of the two components; the emission from the galaxy
progressively becomes less important at wavelengths above $\approx$~4~\micron\ (in the source 
rest frame), where the nuclear reprocessed emission starts emerging significantly 
(see Fig.~\ref{seds}, left panel), being dominant in MIPS at 24~\micron. 
Furthermore, the latter is fully consistent with the upper limits provided in the 
\ks\ band by \cite{mig04}. 

\subsection{Point-like sources}
For the two point-like sources, we adopted a different strategy, since their emission in 
the near-IR is dominated by the unresolved AGN. To reproduce their observed SEDs, 
we extincted a Type~1 quasar template from \cite{elv94} 
with several extinction laws, but we were not able to find a satisfactory solution. 
Then we used the recently published red quasar template from \cite{pol06} 
and found good agreement with the data (Fig.~\ref{seds}, 
right panel), consistently with the results obtained for some obscured AGN in the ELAIS-S1 field 
(\citep{gru07}). 
As in the AGN sub-sample described above, most of the uncertainty lies in the far-IR, where 
a proper study of the SEDs would require MIPS data at 70 and 160~\micron. 

\section{Bolometric corrections}
The determination of the SEDs is meant to be the first step toward the estimate of the bolometric 
luminosities (\lbol) of obscured AGN. The bolometric luminosities can be estimated from the 
luminosity in a given band by applying a suitable bolometric correction \kbol; 
typically, to convert the 2--10~keV luminosity into \lbol, \kbol$\approx$~30 is assumed, 
although this value was derived from the average of few dozens of bright, mostly low-redshift 
Type~1 quasars (\citep{elv94}). 
For obscured sources, only few estimates are present in literature (e.g., \citep{pol06}). 
We derived \kbol\ by integrating the quasar SEDs over the \xray\ 
(0.5--500~keV) and IR (1--1000~\micron) intervals; 
in the \xray\ band, we converted the \hbox{2--10~keV} luminosity assuming a power law with photon 
index $\Gamma=1.9$ (typical for AGN emission) and the observed column density (\citep{per04}). 

To derive the bolometric corrections, we accounted for both the covering factor of the 
absorbing material (i.e., the opening angle of the torus) and the anisotropy of the IR emission. 
According to unification models of AGN, the former effect should be directly related to the 
observed fraction of Type~2/Type~1 AGN which, in the latest models of \cite{gil07}, 
is $\approx$~1.5 in the luminosity range of our sample. 
Furthermore, the torus is likely to re-emit a fraction of the intercepted radiation in a direction 
which does not lie along our line-of-sight; the correction for this anisotropy, according to 
the templates of \cite{sil04}, is $\approx$~10--20\% (given the column densities of our sources). 
Once these corrections are taken into account, we obtain $\langle$\kbol$\rangle\approx35$ 
(median \kbol$\approx$~26), similar to the average value of \cite{elv94}; the Type~1.9 quasar at 
$z$=2.09 has the highest \kbol\ ($\approx$~97); see \cite{poz07} for a discussion on the 
uncertainties in these estimates.

\section{Black hole masses and Eddington ratios}
For the six AGN hosted by elliptical galaxies, we can derive both the galaxy and black hole masses. 
Since the near-IR emission is dominated by the galaxy starlight, we computed 
the rest-frame $L_{\rm K}$ assuming the appropriate SED templates and then the galaxy masses using 
\mstar/$L_{\rm K}\approx0.5-0.9$ (\citep{bc03}); all of our AGN are hosted by massive galaxies 
($\approx1-6\times10^{11}$~\msun).

To estimate the black hole masses, we used the local \mbh--$L_{\rm K}$ (\citep{mh03}) which, along 
with the \mstar/$L_{\rm K}$ values, provides a \mbh--\mstar\ relation. 
Despite several attempts in the recent literature to investigate 
whether and how the black hole mass vs. stellar mass relation 
evolves with cosmic time, there is no consensus yet. In this work, we assume the findings of 
\cite{pen06}, who found that in the redshift range covered by our sources, the \mbh-\mstar\ relation 
evolves by a factor of $\approx$~2 with respect to the local value; see \cite{poz07} for an 
extensive discussion. Under this hypothesis, we obtain black hole masses for the six obscured 
quasars hosted by elliptical galaxies of $\approx2.0\times10^{8}-2.5\times10^{9}$~\msun; 
these values are broadly consistent with the average black hole masses obtained by \cite{md04} 
for the Sloan Digital Sky Survey (SDSS) Type~1 quasars (using optical and ultra-violet mass scaling 
relationships) in our redshift range ($\approx3.5\times10^{8}-8.6\times10^{8}$~\msun). 

As a final step, we derived the Eddington ratios, defined as \lbol/\ledd, where \ledd\ is the 
Eddington luminosity. We note that the uncertainties related to these estimates are clearly large, 
due to the uncertainties of the approach adopted to derive the bolometric luminosities (through the 
templates of \citep{sil04}) and the black hole masses (see above). The average Eddington ratio 
is $\approx$~0.05, suggesting that our obscured quasars may have already passed their 
rapidly accreting phase and are reaching their final masses at low Eddington rates. 
The Eddington ratios of our sources are significantly lower than those derived for the SDSS 
Type~1 quasars in the same redshift range ($\approx$~0.3--0.4, see \citep{md04}).

\section{Summary}
We used optical, near-IR, and \spitzer\ IRAC and MIPS (at 24~\micron) data to unveil the 
reprocessed nuclear emission of eight hard \xray\ selected Type~2 quasars at 
\hbox{$z\approx0.9-2.1$}. 
From proper modelling of the nuclear SEDs, we derived a median (average) bolometric correction of 
$\approx$~26 ($\approx$~35). For the six obscured sources dominated by the host galaxy starlight 
up to near-IR wavelengths, we also derived black hole masses of the order of 
\hbox{$2.0\times10^{8}-2.5\times10^{9}$~\msun} and relatively low Eddington ratios ($\approx$~0.05), 
suggestive of a low-activity accretion phase. 


\begin{theacknowledgments}
The authors acknowledge partial financial support by the Italian Space Agency under the 
contract ASI--INAF I/023/05/0. 

\end{theacknowledgments}





\begin{thebibliography}{0}
\expandafter\ifx\csname natexlab\endcsname\relax\def\natexlab#1{#1}\fi
\providecommand{\enquote}[1]{``#1''}
\expandafter\ifx\csname url\endcsname\relax
  \def\url#1{\texttt{#1}}\fi
\expandafter\ifx\csname urlprefix\endcsname\relax\def\urlprefix{URL }\fi
\providecommand{\eprint}[2][]{\url{#2}}

\end{thebibliography}


\begin{thebibliography}{2}

\bibitem[Giacconi et al. (2002)]{gia02}
R. Giacconi et al., \emph{\apjs} \textbf{139}, 369--410 (2002). 

\bibitem[Alexander et al. (2003)]{ale03}
D.~M. Alexander et al., \emph{\aj} \textbf{126}, 539--574 (2003). 

\bibitem[Fiore et al. (2003)]{fio03}
F. Fiore et al., \emph{\aap} \textbf{409}, 79--90 (2003). 

\bibitem[Brandt \& Hasinger (2005)]{bh05}
W.~N. Brandt and G. Hasinger, \emph{\araa} \textbf{43}, 827--859 (2005). 

\bibitem[Comastri et al. (1995)]{com95}
A. Comastri, G. Setti, G. Zamorani and G. Hasinger, \emph{\aap} \textbf{296}, 1--12 (1995). 

\bibitem[Gilli et al. (2007)]{gil07}
R. Gilli, A. Comastri and G. Hasinger, \emph{\aap} \textbf{463}, 79--96 (2007). 

\bibitem[Bauer et al. (2004)]{feb04}
F.~E. Bauer, D.~M. Alexander, W.~N. Brandt, D.~P. Schneider, E. Treister, A.~E. Hornschemeier 
and G.~P. Garmire, \emph{\aj} \textbf{128}, 2048--2065 (2004).

\bibitem[Hickox \& Markevitch (2006)]{hm06}
R.~C. Hickox and M. Markevitch, \emph{\apj} \textbf{645}, 95--114 (2006). 

\bibitem[Barger et al. (2005)]{bar05}
A.~J. Barger, L.~L. Cowie, R.~F. Mushotzky, Y. Yang, W.-H. Wang, A.~T. Steffen and P. Capak, 
\emph{\aj} \textbf{129}, 578--609 (2005). 

\bibitem[Hopkins et al. (2006)]{hop06}
P.~F. Hopkins, L. Hernquist, T.~J. Cox, T. Di Matteo, B. Robertson and V. Springel, 
\emph{\apjs} \textbf{163}, 1--49 (2006). 

\bibitem[Elvis et al. (1994)]{elv94}
M. Elvis et al., \emph{\apjs} \textbf{95}, 1--68 (1994). 

\bibitem[Richards et al. (2006)]{ric06}
G.~T. Richards et al., \emph{\apjs} \textbf{166}, 470--497 (2006).

\bibitem[Polletta et al. (2006)]{pol06}
M. Polletta et al., \emph{\apj} \textbf{642}, 673--693 (2006). 

\bibitem[Baldi et al. (2002)]{bal02}
A. Baldi, S. Molendi, A. Comastri, F. Fiore, G. Matt and C. Vignali, \emph{\apj} \textbf{564}, 
190--195 (2002). 

\bibitem[Cocchia et al. (2007)]{coc07}
F. Cocchia et al., \emph{\aap, in press}, \eprint{astro-ph/0612023} (2007). 

\bibitem[Mignoli et al. (2004)]{mig04}
M. Mignoli et al., \emph{\aap} \textbf{418}, 827--840 (2004).

\bibitem[Maiolino et al. (2006)]{mai06}
R. Maiolino et al., \emph{\aap} \textbf{445}, 457--463 (2006). 

\bibitem[Vignali et al. (2006)]{vig06}
C. Vignali, D.~M. Alexander and A. Comastri, \emph{\mnras} \textbf{373}, 321--329 (2006). 

\bibitem[Pozzi et al. (2007)]{poz07}
F. Pozzi et al., \emph{\aap, in press}, \eprint{arXiv:0704.0735} (2007). 


\bibitem[Bruzual \& Charlot (2003)]{bc03}
G. Bruzual and S. Charlot, \emph{\mnras} \textbf{344}, 1000--1028 (2003).

\bibitem[Silva et al. (2004)]{sil04}
L. Silva, R. Maiolino and G.~L. Granato, \emph{\mnras} \textbf{355}, 973--985 (2004). 

\bibitem[Granato \& Danese (1994)]{gd94}
G.~L. Granato and L. Danese, \emph{\mnras} \textbf{268}, 235--252 (1994). 

\bibitem[Perola et al. (2004)]{per04}
G.~C. Perola et al., \emph{\aap} \textbf{421}, 491--501 (2004). 

\bibitem[Gruppioni et al. (2007)]{gru07}
C. Gruppioni et al., \emph{\aap, in preparation}. 



\bibitem[Marconi \& Hunt (2003)]{mh03}
A. Marconi and L.~K. Hunt, \emph{\apjl} \textbf{589}, L75--L77 (2003). 

\bibitem[Peng et al. (2006)]{pen06}
C.~Y. Peng, C.~D. Impey, L.C. Ho, E.~J. Barton and H.-W. Rix, \emph{\apj} \textbf{640}, 114--125 
(2006). 

\bibitem[McLure \& Dunlop (2004)]{md04}
R.~J. McLure and J.~S. Dunlop, \emph{\mnras} \textbf{352}, 1390--1404 (2004). 

\end{thebibliography}

\IfFileExists{\jobname.bbl}{}
 {\typeout{}
  \typeout{******************************************}
  \typeout{** Please run "bibtex \jobname" to optain}
  \typeout{** the bibliography and then re-run LaTeX}
  \typeout{** twice to fix the references!}
  \typeout{******************************************}
  \typeout{}
 }



\end{document}